\documentclass[journal]{IEEEtran}
\usepackage{amsmath}
\usepackage{caption}

\usepackage[pdftex]{graphicx}
\graphicspath{{../pdf/}{../jpeg/}}

\usepackage[pdftex]{graphicx}

\usepackage{array}
\usepackage{mdwmath}
\usepackage{mdwtab}
\usepackage{eqparbox}
\usepackage{url}

\usepackage{booktabs}

\begin{document}
\bstctlcite{IEEEexample:BSTcontrol}
    \title{Adaptive Multiscale Retinal Diagnosis: A Hybrid Trio-Model Approach for Comprehensive Fundus Multi-Disease Detection Leveraging Transfer Learning and Siamese Networks}
  \author{Yavuz Selim Inan\IEEEmembership{}\\
      ARSEF Grand Award Winner Research

}

\maketitle

\begin{abstract}

WHO has declared that more than 2.2 billion people worldwide are suffering from visual disorders, such as media haze, glaucoma, and drusen. At least 1 billion of these cases could have been either prevented or successfully treated, yet they remain unaddressed due to poverty, a lack of specialists, inaccurate ocular fundus diagnoses by ophthalmologists, or the presence of a rare disease. To address this, the research has developed the Hybrid Trio-Network Model Algorithm for accurately diagnosing 12 distinct common and rare eye diseases. This algorithm utilized the RFMiD dataset of 3,200 fundus images and the Binary Relevance Method to detect diseases separately, ensuring expandability and avoiding incorrect correlations. Each detector, incorporating finely tuned hyperparameters to optimize performance, consisted of three feature components: A classical transfer learning CNN model, a two-stage CNN model, and a Siamese Network. The diagnosis was made using features extracted through this Trio-Model with Ensembled Machine Learning algorithms. The proposed model achieved an average accuracy of 97\% and an AUC score of 0.96. Compared to past benchmark studies, an increase of over 10\% in the F1-score was observed for most diseases. Furthermore, using the Siamese Network, the model successfully made predictions in diseases like optic disc pallor, which past studies failed to predict due to low confidence. This diagnostic tool presents a stable, adaptive, cost-effective, efficient, accessible, and fast solution for globalizing early detection of both common and rare diseases.
\end{abstract}


\IEEEpeerreviewmaketitle

\section{Introduction}

\IEEEPARstart {M}{ore} than 2.2 billion people worldwide are suffering from visual disorders [1], such as
media haze, diabetic retinopathy, glaucoma, central serous retinopathy, epiretinal membrane,
drusen, branch retinal vein occlusion, myopia, tessellation, optic disc pallor, refractive errors,
presbyopia, age-related macular degeneration. The World Health Organization [WHO] has
declared that at least 1 billion of these cases could have been either prevented or treated
successfully, yet they remain unaddressed [1]. Timely and accurate diagnosis and treatment are
crucial, as these visual disorders can lead to permanent impairment and even preventable
blindness [2; 3; 4]. However, due to various reasons such as poverty, costly diagnostic
radiographic tests, or lack of specialists in their area, millions of people lack access to
ophthalmic services for diagnosis. Non-eye care providers often struggle with accurate ocular
fundus examinations but employing Artificial Intelligence [AI] could enable them to make
correct referrals despite these challenges [4]. Even without these problems, patients may still
require a referral for a second and credible opinion to receive appropriate treatment with
concerns about the credibility of their diagnoses. Ophthalmologists also need the perspective of
another professional due to the frequent diagnostic errors and high costs of claims associated
with those errors. Research in ophthalmology has revealed significant diagnostic errors,
identifying that both rare and some common diseases, which may seem obvious, are often missed
by the human eye [4, 5, 6, 7, 8]. This study aims to achieve early and accurate detection of
common and rare diseases by developing The Hybrid Trio-Network Model Algorithm, which is
designed to be adaptive and expandable for future integration of new disease detectors,
benefiting both patients and ophthalmologists globally.\\

\begin{table*}[ht]
\centering
\caption{RFMiD Data [14] Table created by the Author}
\label{tab:rfmid}
\begin{tabular}{|c|l|c|c|c|c|}
\hline
\textbf{Acronym} & \textbf{Name} & \textbf{Training} & \textbf{Validation} & \textbf{Testing} & \textbf{Total} \\
\hline
N & Normal & 1519 & 506 & 506 & 2531 \\
\hline
DN & Drusen & 138 & 46 & 46 & 230 \\
\hline
ODC & Optic Disc Cupping & 282 & 72 & 91 & 445 \\
\hline
TSLN & Tessellation & 186 & 65 & 53 & 304 \\
\hline
ARMD & Age-Related Macular Degeneration & 100 & 38 & 31 & 169 \\
\hline
RS & Retinitis & 43 & 14 & 14 & 71 \\
\hline
ODE & Optic Disc Edema & 58 & 21 & 17 & 96 \\
\hline
ODP & Optic Disc Pallor & 65 & 26 & 24 & 115 \\
\hline
DR & Diabetic Retinopathy & 376 & 132 & 124 & 632 \\
\hline
MH & Media Haze & 317 & 102 & 104 & 523 \\
\hline
BRVO & Branch Retinal Vein Occlusion & 73 & 23 & 23 & 119 \\
\hline
MYA & Myopia & 101 & 34 & 32 & 167 \\
\hline
CRVO & Central Retinal Vein Occlusion & 28 & 8 & 9 & 45 \\
\hline
\textbf{Total} & & \textbf{3286} & \textbf{1087} & \textbf{1074} & \textbf{5447} \\
\hline
\end{tabular}
\end{table*}
At this juncture, the emerging use of AI for diagnosing through fundus photos is becoming increasingly popular, as it enables ophthalmologists to verify the presence of diseases
that would otherwise be missed. AI can offer an accessible, flexible, affordable, fast, and
credible solution for life-saving eye care. Capturing a color fundus photograph with a
professional camera, a smartphone, or a telemedicine device and then analyzing the image using
AI for diagnostic purposes will make the diagnostic process very affordable and accessible.
Additionally, the color photograph and AI results can be sent electronically to ophthalmologists
or eye care services for further evaluation, a second opinion, or ongoing follow-up. This
approach streamlines the process of eye disease diagnosis and management. There is a growing
body of literature on the application of AI showing that the integration of AI in ophthalmology
aims to fill the diagnostic accuracy gap, with the potential to significantly improve eye care (e.g.,
2; 4; 9; 10; 11; 12; 13). However, there is still much to be explored or developed regarding AI,
including areas not yet addressed or lacking in AI diagnostics, especially the importance of
creating an AI that is trustworthy in detecting and diagnosing fundus diseases, both
common and rare. \\

While numerous studies have been conducted in the field of ophthalmology to detect
retinal diseases with AI models, a significant challenge remains unaddressed: The imbalance in
disease prevalence and data availability limitations. Existing research has often struggled in
scenarios where data is extremely limited, particularly in the case of rare ocular fundus diseases.
Therefore, there is still much to explore or develop in AI diagnostics, particularly in creating an
AI system that is highly trustworthy in detecting and diagnosing fundus diseases, both common
and rare. \\

The significance of the current study lies in the development, training, validation, and
testing of a stable and adaptive AI model. This model is designed to automatically diagnose
multiple different common and rare eye diseases from retinal color photographs with high
accuracy. The study’s potential impact on society is significant. In this study, the fully automated
robust model, utilizing hybrid algorithms, aimed to provide a faster, more accessible, costeffective, and accurate diagnosis, thereby benefiting patients and ophthalmologists. More
specifically, it aimed to safeguard ophthalmologists against the high costs associated with
frequent diagnostic errors, which are common in rare diseases and may often go unnoticed by the
human eye. \\

\section{METHODOLOGY}

\subsection{Data and Preprocessing}

This research utilized a publicly available, de-identified dataset of 3,200 fundus images
named “Retinal Fundus Multi-Disease Image Dataset” [RFMiD] [14] for the training, validation,
and testing of the algorithm. These 3,200 fundus images were captured using three distinct types
of digital fundus cameras, namely TOPCON 3D OCT-2000, Kowa VX-10$\alpha$, and TOPCON TRCNW300. They were centered either on the optic disc or macula. Those images have been
annotated for multiple different conditions, based on the adjudicated consensus reached by two
experienced senior retinal specialists [14]. Pachade et al (2021) states, "The disease wise
stratification on average in training, evaluation and test set is 60 ± 7 \%, 20 ± 7\%, and 20 ± 5\%,
respectively” [14]. The current study used 1920 images (60\%) in the training set, 640 images
(20\%) in the validation set, and 640 images (\%20) in the testing set. Each set will include a
balanced representation of various disease probabilities, ensuring a comprehensive and
representative dataset for model training, validation, and testing (see Table 1). \\

Image processing (see Figure 1) included cropping and histogram balancing. Then, for
each disease, data were balanced using oversampling methods, and images were subjected to
augmentation techniques such as rotation, random crop, shear, Gaussian noise, pixel noise, blur,
zoom, flip, and brightness change. Following that, for component 2, Sobel Edge Detection,
Posterize Effect, and Emboss Effect were applied. \\

\begin{figure}[h]
\centering
\includegraphics[width=\columnwidth]{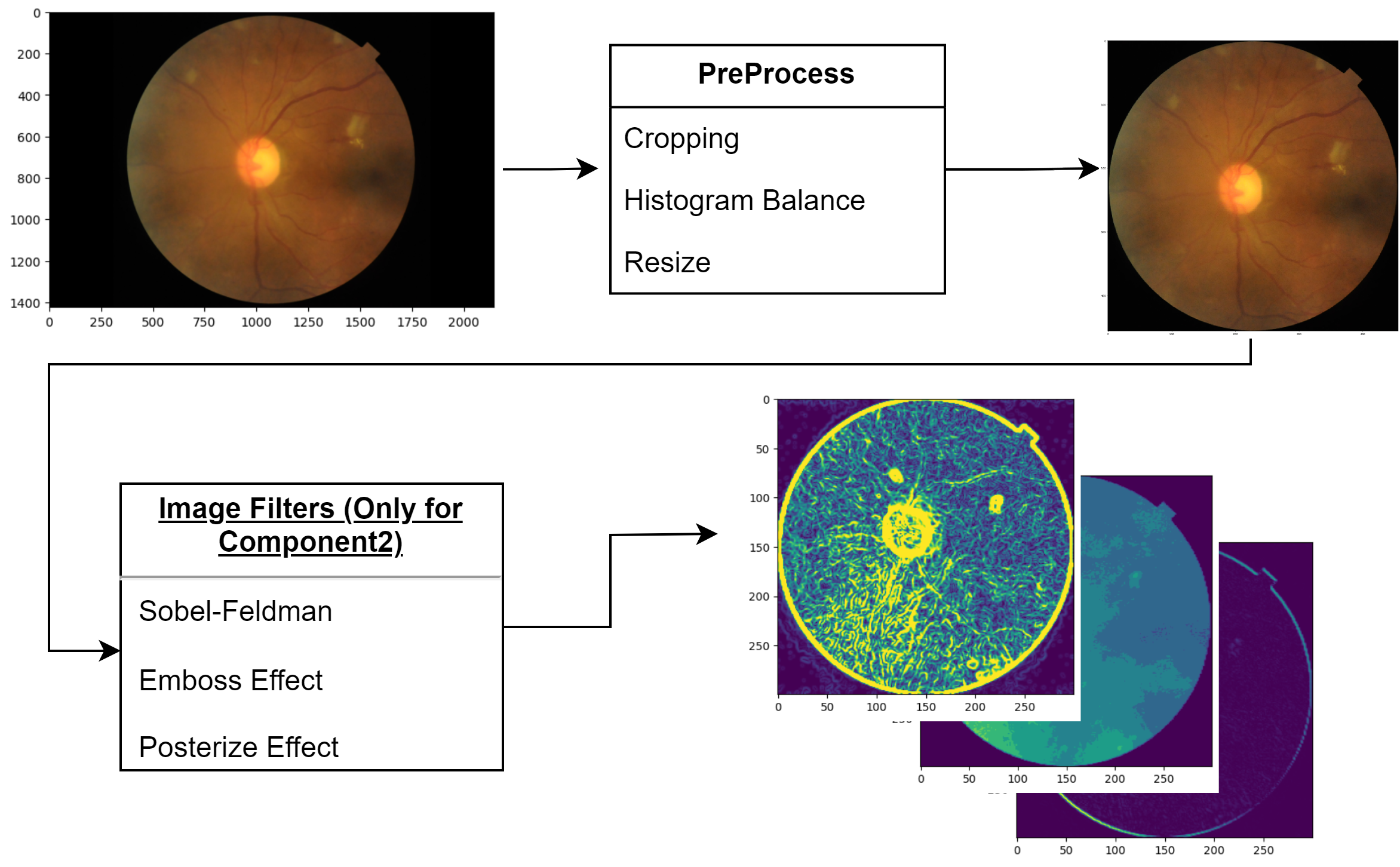}
\caption{Reprocessing Graphics created by the Author except the Retinal Image from Data [14]}
\label{fig1}
\end{figure}

\begin{figure*}[h]
\centering
\includegraphics[width=2\columnwidth]{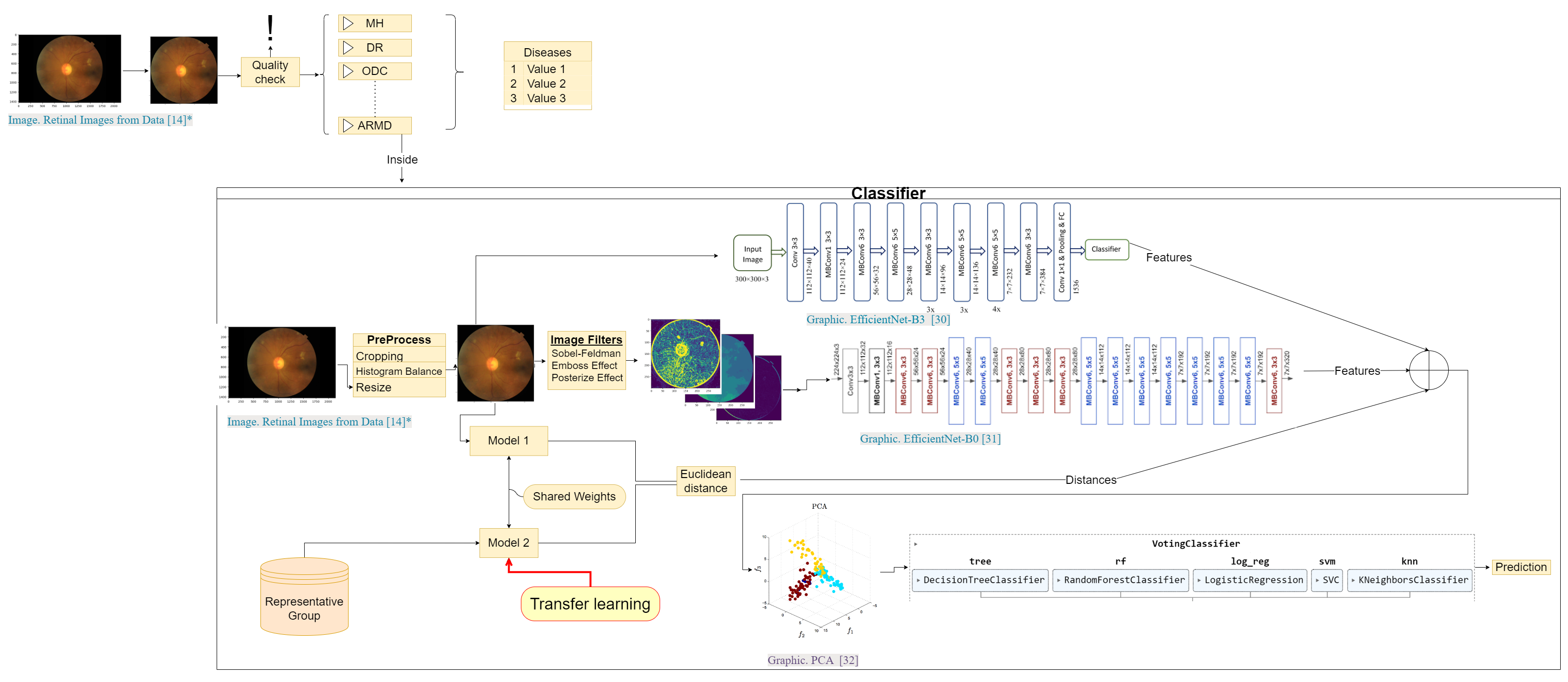}
\caption{Image of workflow created by the Author except graphics by the retinal image from
data [14], schematic representation of EfficientNet-B3 [30], EfficientNet-B0 [31], and the
PCA graphic (Graphic from D'Aspremont et al., 2007, data from Iconix Pharmaceuticals, reproduced by and as cited in [32]) (each explained below).}
\label{fig1}
\end{figure*}

\subsection{Model}
The current study employed the Binary Relevance Method [15; 16] to establish a multidisease classification system. This system enabled the algorithm to detect each disease
separately, ensuring expandability and avoiding incorrect correlations between diseases in the
dataset. This automatic multi-disease detection model created multiple independent disease
detectors, one for each disease, allowing independent detection of multiple diseases in a single
diagnostic process. Each disease detector, incorporating finely tuned hyperparameters to
optimize performance, consists of three feature components: \\

\begin{enumerate}
  \item A classical transfer learning Convolutional Neural Network [CNN] model for complex
feature extraction [17; 18; 30]

  \item A two-stage CNN model for advanced processing and refinement of these features, that
uses Sobel Edge Detection, Posterize Effect, and Emboss Effect [19; 20; 21; 31]
  \item A Siamese Network for enhanced comparative analysis [22; 23; 33]. Subsequently,
features extracted through this Trio-Network Model approach underwent Principal
Component Analysis [PCA] [24; 32] and predictions were made using voting classifiers
and machine learning algorithms (see Figure 2) \\
\end{enumerate}

\subsubsection{Component 1}

The first component of the model (see Figure 3) utilizes transfer learning via
EfficientNetB3 [30]. This approach is fine-tuned by adding two layers whose dimensions and
properties are fine-tuned during hyperparameter optimization of the pre-trained network for more
nuanced feature extraction. CNN is pre-trained on the ImageNet dataset. This method is
particularly effective in leveraging learned patterns from extensive datasets, thus reducing the
need for large-scale data in medical imaging. The binary cross-entropy loss function is used, as a
training loss function. [26; 27] 

\begin{figure}[h]
\centering
\includegraphics[width=\columnwidth]{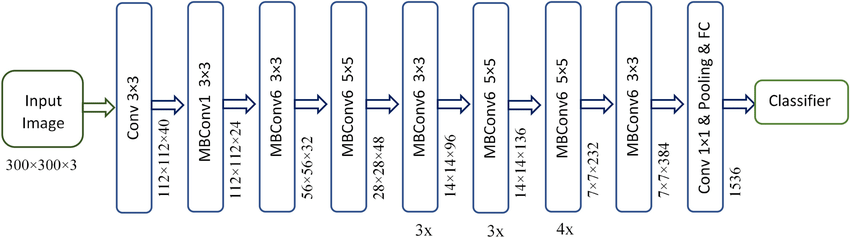}
\caption{Schematic Representation of EfficientNet-B3 [30]}
\label{fig3}
\end{figure}

\begin{equation}
L_{BCE} = -\frac{1}{N} \sum_{i=1}^{N} \left[ y_i \cdot \log(\hat{y}_i) + (1 - y_i) \log(1 - \hat{y}_i) \right]
\end{equation}
Formula 1: Binary Cross Entropy Loss

\subsubsection{Component 2}
The second component (see Figure 4) combines manual feature extraction methods with
CNNs via EfficientNet-B0 [31]. This hybrid approach broadens the range of detectable features,
making it invaluable, especially in scenarios with limited data. The component starts with edge
detection using Sobel filters, emboss, and posterize effects. The Sobel operator calculates the
gradient of the image intensity, effectively highlighting regions with high spatial frequency that
correspond to edges. And the posterize effect reduces the number of color levels in an image.
Each pixel value is quantized to the nearest value within a set range, reducing the image's overall
color complexity. And the emboss effect highlights edges by replacing each pixel with a
highlight or shadow. After the initial feature combination in Stage 1, these features are further
processed using the EfficientNet-B0 architecture, a deep Convolutional Neural Network. As a
training loss function, the binary cross-entropy loss function is utilized.

\begin{figure}[h]
\centering
\includegraphics[width=0.9\columnwidth]{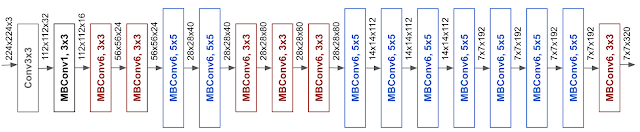}
\caption{Schematic Representation of EfficientNet-B0 [31]}
\label{fig4}
\end{figure}

\subsubsection{Component 3}
The third component of the model (see Figure 5) leverages a Siamese network
architecture [33], which plays a pivotal role in the few-shot learning approach, especially useful
in scenarios with extremely limited data or for very rare diseases. Siamese networks focus on
learning a similarity metric between input pairs, aiming to minimize the distance between similar
items and maximize the distance between dissimilar ones. This approach is part of meta-learning,
where the network learns to learn from small data sets, making it highly effective for few-shot
learning tasks.

\begin{figure}[h]
\centering
\includegraphics[width=\columnwidth]{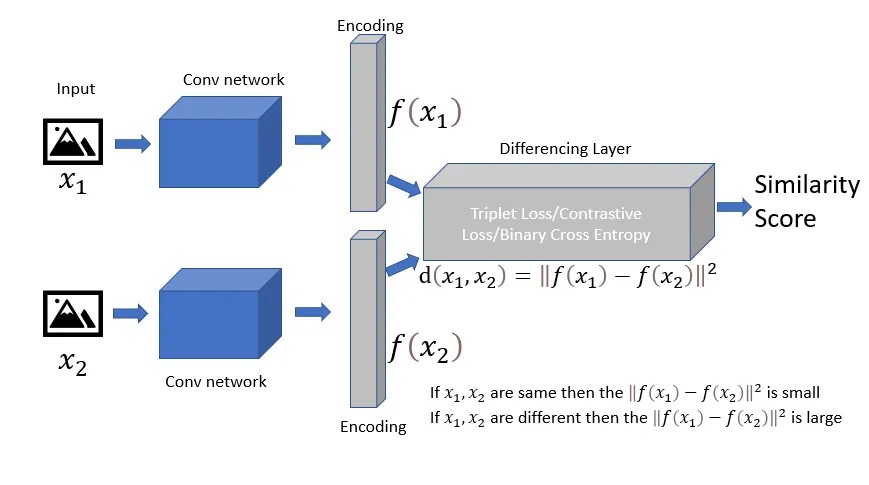}
\caption{Siamese Network [33]}
\label{fig5}
\end{figure}

\subsubsection{Machine Learning}
Subsequently, the features extracted using the Trio-Network Model approach underwent
Principal Component Analysis (PCA) [24; 25; 32]. Predictions are then made using voting
classifiers and a variety of machine-learning algorithms (see fig. 6). \\

These features are processed with an ensemble of machine learning algorithms, including decision tree classifiers, random forest classifiers, logistic regression, support vector machines, and k-nearest
neighbors. This comprehensive strategy ensures robust performance even in scenarios where data
distribution is skewed, making it a powerful solution to the issue of data imbalance. \\

\begin{figure}[h]
\centering
\includegraphics[width=\columnwidth]{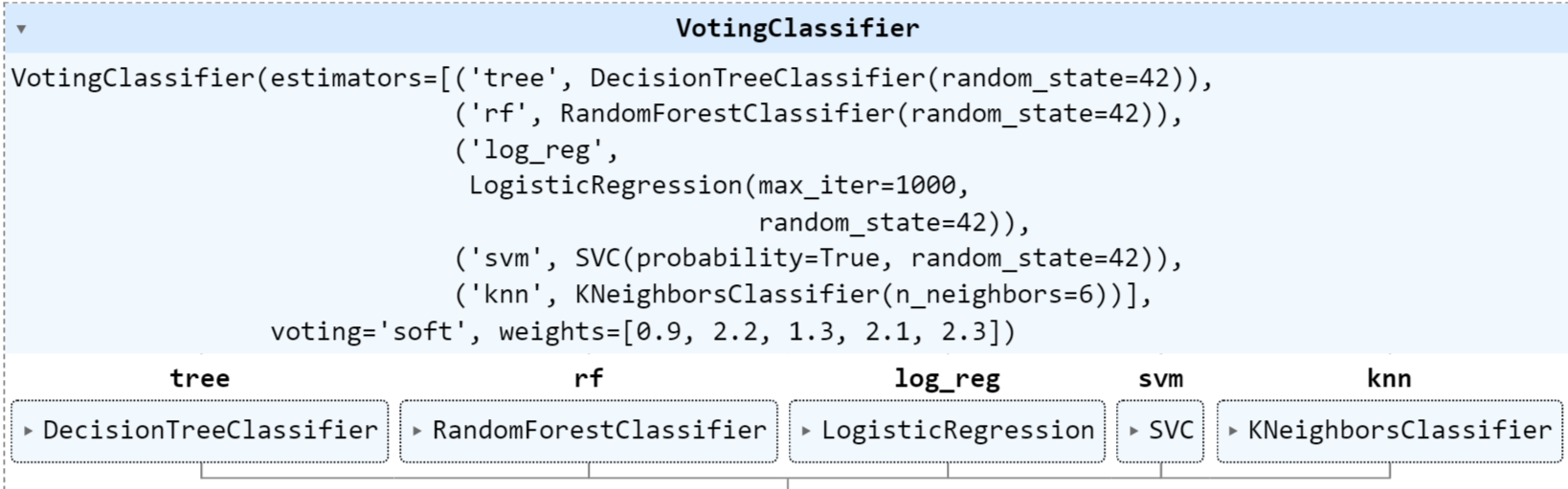}
\caption{Diagram by the Author}
\label{fig6}
\end{figure}

\section{RESULTS}
These graphics (see Figure 7, 8, 9, 10) display ROC, which are Receiver Operating
Characteristic Curves, based on the testing set (n=640) \\ \\ \\ \\

\begin{figure}[h]
\centering
\begin{minipage}[b]{0.48\columnwidth}
    \centering
    \includegraphics[width=\linewidth]{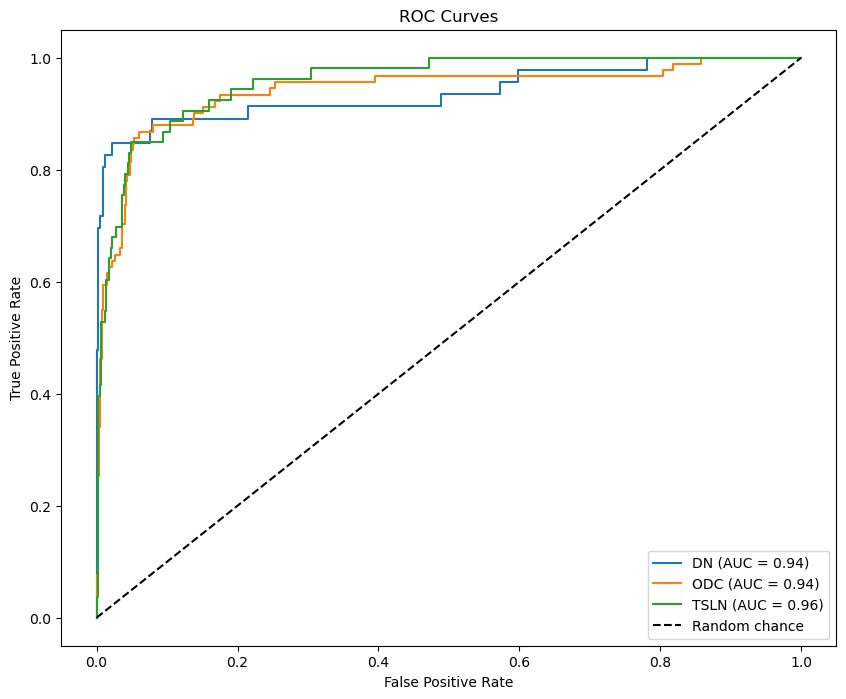}
    \caption{Receiver Operating Characteristic Curve Graphics by the Author }
    \label{fig:image1}
\end{minipage}
\hfill
\begin{minipage}[b]{0.48\columnwidth}
    \centering
    \includegraphics[width=\linewidth]{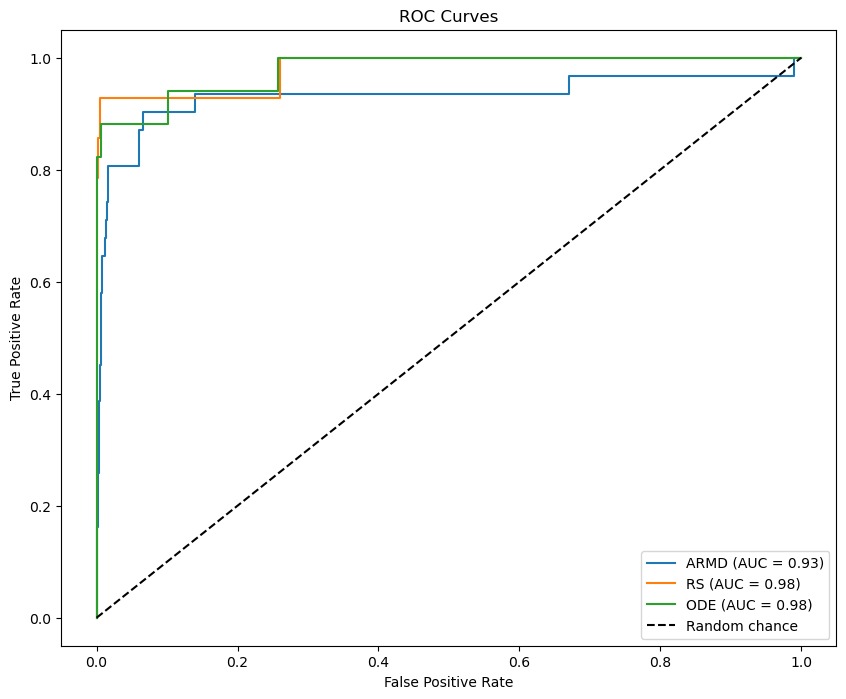}
    \caption{Receiver Operating Characteristic Curve Graphics by the Author }
    \label{fig:image2}
\end{minipage}
\end{figure}

\begin{figure}[h]
\centering
\begin{minipage}[b]{0.48\columnwidth}
    \centering
    \includegraphics[width=\linewidth]{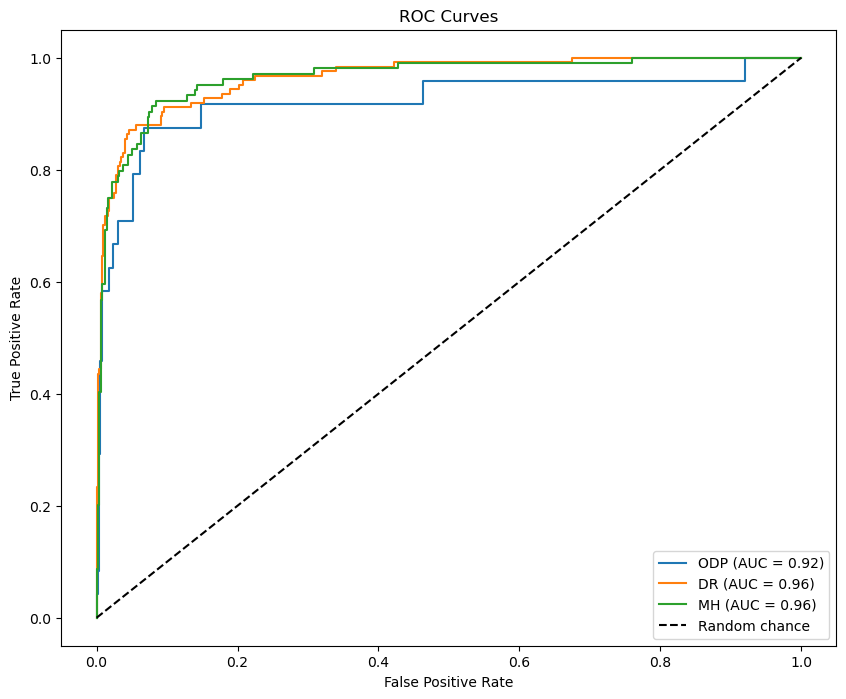}
    \caption{Receiver Operating Characteristic Curve Graphics by the Author}
    \label{fig:image1}
\end{minipage}
\hfill
\begin{minipage}[b]{0.48\columnwidth}
    \centering
    \includegraphics[width=\linewidth]{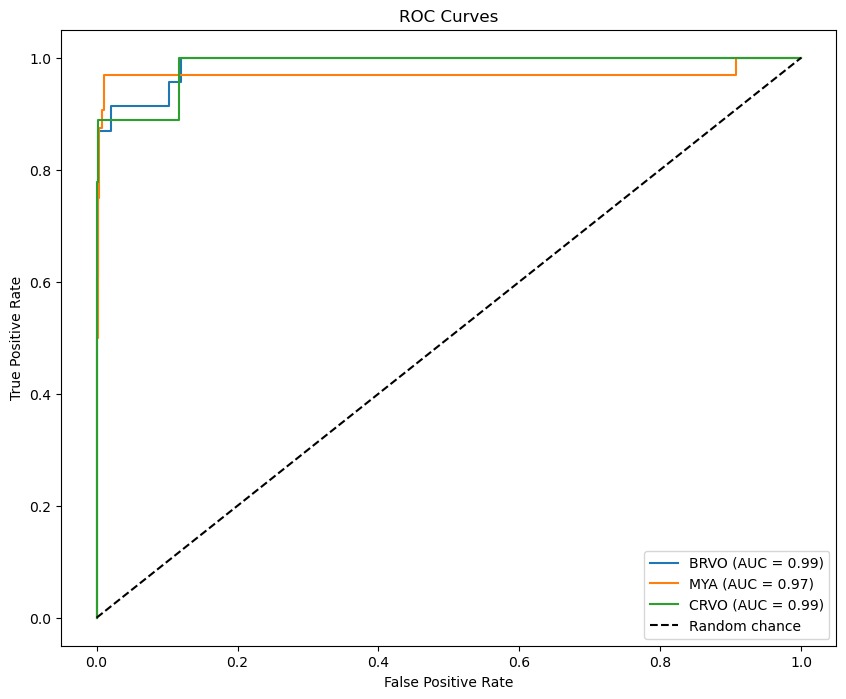}
    \caption{Receiver Operating Characteristic Curve Graphics by the Author}
    \label{fig:image2}
\end{minipage}
\end{figure}

Table 2 provides a concise comparison of the performance metrics—Accuracy, Precision,
Recall, F1 score, and Area Under the Curve [AUC]—for various eye diseases analyzed by the
Hybrid Trio-Network Model Algorithm. The average metrics across all diseases highlight the
algorithm's effectiveness in diagnosing both common and rare diseases from retinal
photographs.

\begin{table}[h]
\centering
\caption{Performance Metrics of the Hybrid Trio-Network Model Algorithm}
\label{tab:performance}
\begin{tabular}{|l|c|c|c|c|c|}
\hline
\textbf{Disease} & \textbf{Accuracy} & \textbf{Precision} & \textbf{Recall} & \textbf{F1} & \textbf{AUC} \\
\hline
DN & 0.97344 & 0.89189 & 0.71739 & 0.79518 & 0.93698 \\
\hline
ODC & 0.92500 & 0.78667 & 0.64835 & 0.71084 & 0.94069 \\
\hline
TSLN & 0.95469 & 0.87500 & 0.52830 & 0.65882 & 0.95789 \\
\hline
ARMD & 0.97188 & 0.74074 & 0.64516 & 0.68966 & 0.93130 \\
\hline
RS & 0.99375 & 0.91667 & 0.78571 & 0.84615 & 0.98095 \\
\hline
ODE & 0.99531 & 1.00000 & 0.82353 & 0.90323 & 0.97857 \\
\hline
ODP & 0.97188 & 0.68750 & 0.45833 & 0.55000 & 0.92100 \\
\hline
DR & 0.93750 & 0.86207 & 0.80645 & 0.83333 & 0.96488 \\
\hline
MH & 0.93906 & 0.82828 & 0.78846 & 0.80788 & 0.96466 \\
\hline
BRVO & 0.99219 & 0.90909 & 0.86957 & 0.88889 & 0.98901 \\
\hline
MYA & 0.99063 & 0.93333 & 0.87500 & 0.90323 & 0.96993 \\
\hline
CRVO & 0.99688 & 1.00000 & 0.77778 & 0.87500 & 0.98679 \\
\hline
\textbf{Average} & \textbf{0.97018} & \textbf{0.86927} & \textbf{0.72700} & \textbf{0.78852} & \textbf{0.96022} \\
\hline
\end{tabular}
\end{table}

The analysis of various models reveals notable results: An impressive average accuracy
of 97.02\%, with the highest accuracy reported for ODE at 99.531\%, and the lowest for ODC at
92.5\%. Precision across models averages at 86.927\%, with ODE achieving perfect precision at
100\%, and ODP at the lower end with 68.75\%. The models demonstrate an ability to identify
actual positives, evidenced by an average recall of 72.7\%, where ODE again leads with
82.3529\%, and TSLN has the lowest at 52.8302\%. The balance between precision and recall is
quantified by an average F1 score of 78.8518\%, with ODE recording the highest at 90.3226\%
and ODP the lowest at 55\%.
Furthermore, the models exhibit strong separability between classes, as shown by an
average AUC score of 0.96, with RS achieving the highest at 0.980945 and ODP the lowest at
0.920996. These numbers reflect the models' exceptional capability in making accurate
predictions, with their high performance across accuracy, precision, recall, F1 score, and AUC
scores confirming their effectiveness and reliability in their respective tasks.
\section{DISCUSSION}
Figure 11 highlights the comparison of Computed F1 scores against Benchmark F1 [29]
scores for various eye diseases, along with the percentage difference, showcasing the
effectiveness of the Hybrid Trio-Network Model Algorithm in diagnosing both common and rare
diseases from retinal photographs.

\begin{figure}[h]
\centering
\includegraphics[width=\columnwidth]{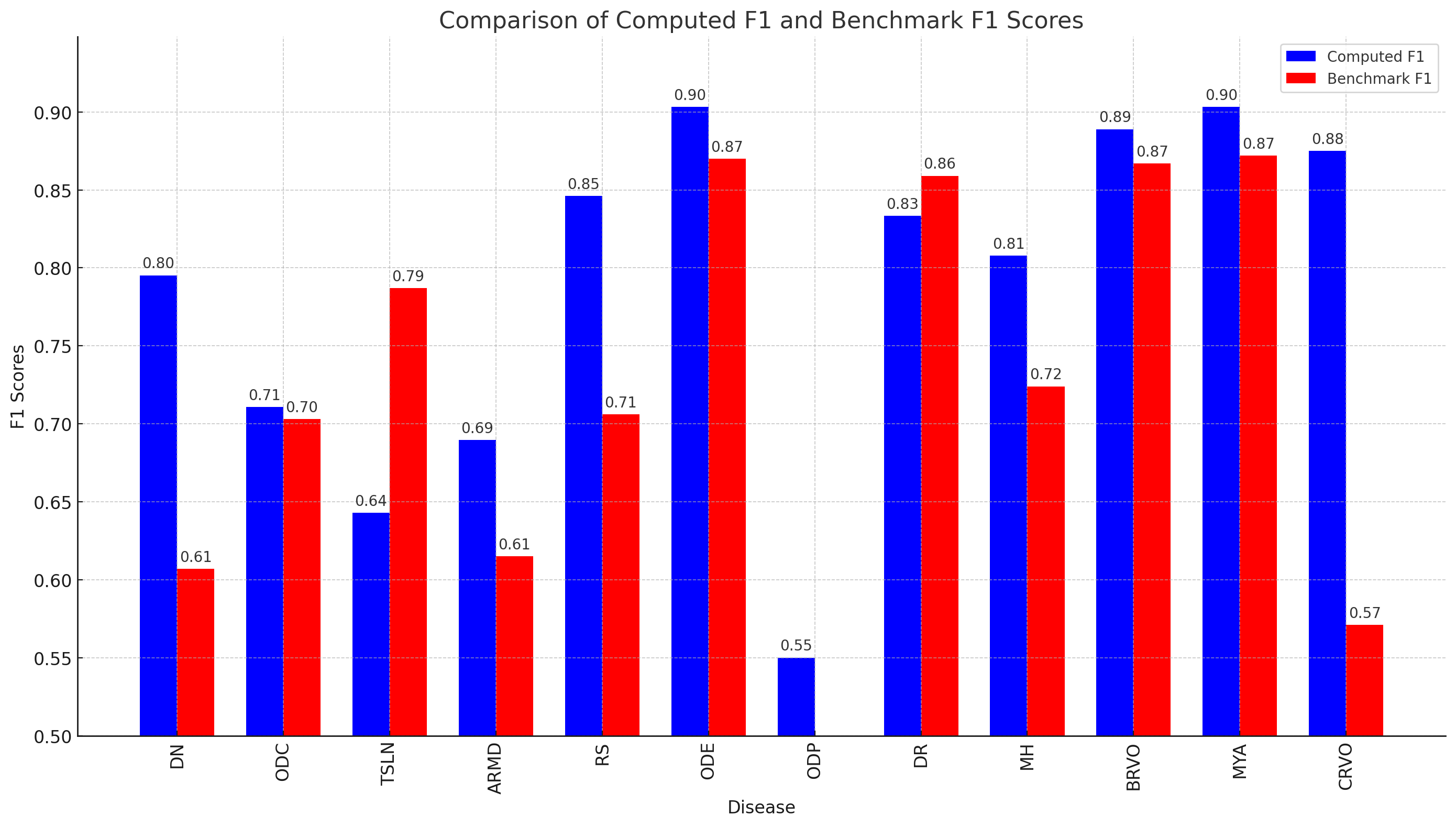}
\caption{Graphics by the Author comparing Computed F1 in the Current Study and Benchmark
F1 Scores [29]}
\label{fig10}
\end{figure}

Compared to past benchmark studies, an increase of over 10\% in the F1-scores was
observed for most diseases, with remarkable improvements for specific conditions. For instance,
the F1-score for diseases like DN (Diabetic Neuropathy) saw an increase from a benchmark of
0.607 to 0.795, marking a 31\% improvement. Similarly, ARMD (Age-Related Macular
Degeneration) and RS (Retinoschisis) experienced an uplift of 12.14\% and 19.85\%, respectively,
compared to their benchmarks.

Furthermore, using the Siamese Network, the model successfully made predictions in
diseases like Optic Disc Pallor [ODP], which past studies failed to predict due to low confidence,
achieving a groundbreaking improvement from 0 to an F1-score of 0.55. Central Retinal Vein
Occlusion [CRVO] demonstrated the highest percentage difference, with an F1-score
improvement of 53.24\% over its benchmark. Future research could extend performance
evaluation beyond accuracy, F-scores, and ROC curves [28].

\section{CONCLUSION}
In conclusion, this research introduces a Hybrid Trio-Network Model Algorithm,
leveraging the Binary Relevance Method using fundus images to diagnose 12 common and rare
eye diseases accurately, namely: Drusen, Optic Disc Cupping, Tessellation, Age-Related Macular
Degeneration, Retinitis, Optic Disc Edema, Diabetic Retinopathy, Media Haze, Branch Retinal
Vein Occlusion, Myopia, and even Optic Disc Pallor, which past studies failed to predict due to
low confidence. The results of the data support the hypothesis. Therefore, it can be concluded
that the Hybrid Trio-Network Model Algorithm, an adaptive, multiscale retinal diagnostic tool,
successfully demonstrates high accuracy in diagnosing both common and rare diseases from
retinal photographs. This approach combined classical transfer learning CNN models, two-stage
CNN models, and Siamese Networks, achieving high diagnostic performance with an average
accuracy of 97.02\%, an F1 score of 78.8518\%, and an AUC score of 0.96. The model
significantly surpassed past benchmarks, demonstrating over a 10\% increase in the F1-score for
most diseases and notable improvements in conditions previously challenging to diagnose. The
diagnostic tool presented a stable, adaptive, efficient, accessible, and fast solution for globalizing
early detection of both common and rare diseases.

\begin{IEEEbiography}[{\includegraphics[width=1in,height=1.25in,clip,keepaspectratio]{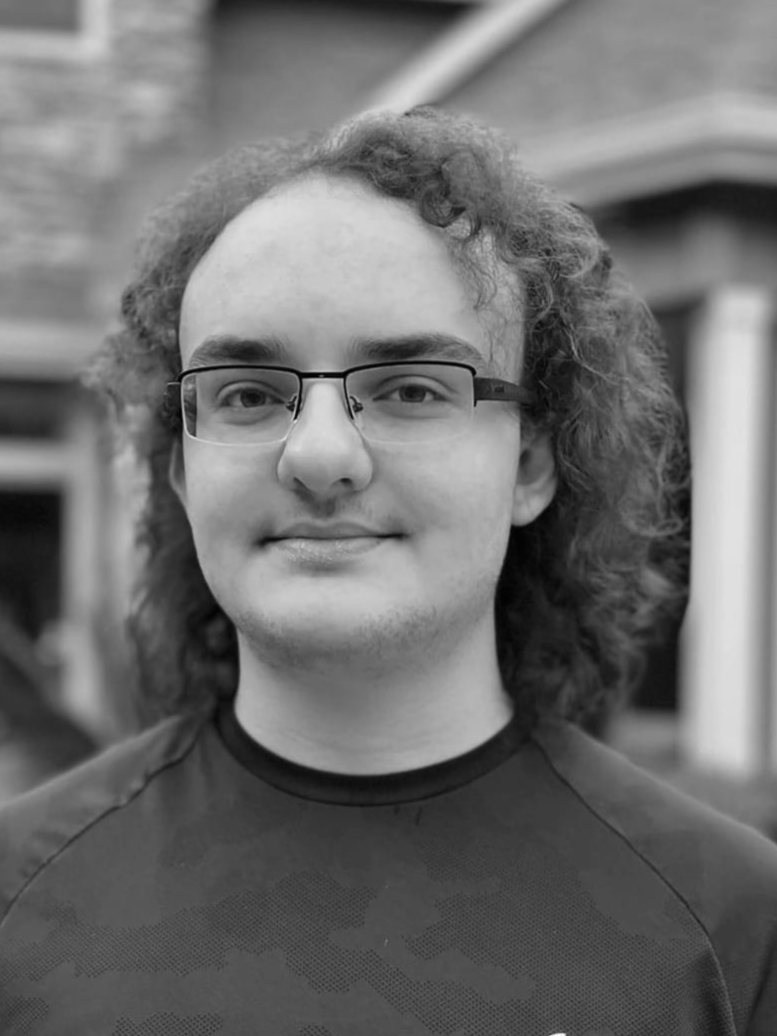}}]{Yavuz Selim Inan}
is a high school student at Ronald W. Reagan High School, San Antonio, TX, expected to graduate in 2026. He has interned at Texas A\&M University's Integrated Biomedical Sensing and Imaging Laboratory, focusing on bio-indicators and image processing algorithms. He developed AI models for retinal disease detection, recognized at ARSEF and TXSEF. He also interned at uVox INC and Kahve Game Studio, working on mobile applications and VR simulations. Yavuz was a member of the Turkish Junior National Computer Olympiad team and represented Turkey at the European Junior Olympiad in Informatics (EJOI). His interests include imaging and sensing technologies, as well as high-efficiency systems.
\end{IEEEbiography}

\vfill

\end{document}